# Third Harmonic Upconverted Full-Stokes Imaging with High-Efficiency Germanium Metasurface from MWIR to SWIR.

HOSNA SULTANA

*School of Electrical & Computer Engineering, University of Oklahoma, 110 W Boyd St, Norman, OK 73019, USA*
*hosna@ou.edu*

**Abstract:** Out of the several optical nonlinear interactions, higher-order harmonic generation is gaining much interest in the context of the resonant metasurface-mediated optical nonlinearity. Instead of a bulky nonlinear crystal as a medium, the thin dielectric resonant metasurface can be utilized for efficient high harmonic generation, which has been explored here in this effort for a high-efficient Germanium metasurface for upconverted full-Stokes imaging at 1.33-micron wavelength by the third harmonic generation (THG) from the 4-micron fundamental wavelength. The nonlinear metasurface of the cross-triangle shape Ge-nanoantenna of aspect ratio of 7.16-1.45 range with FDTD simulation and quantitatively analyzed TH response. The internal TH conversion efficiency for the Ge film is about $1\times10^{-5}$ % to $5.6\times10^{-3}$ % and for the Ge-metasurface $1.5\times10^{-4}$ % to $1.05\times10^{-1}$ % for linearly polarized (LP) incidence, and $1.3\times10^{-4}$ % to $6.1\times10^{-2}$ % for circularly polarized (CP) incidence, for the optical intensity range 0.47 GW/cm$^2$ to 16.8 GW/cm$^2$ respectively. The effect of the Ge film height variation has been discussed and compared with transmission line theory for the nonlinear medium. The metasurface design pitch is higher than the third-harmonic wavelength to enable multimodal TH diffraction orders, which is suitably tuned by the anisotropic cross-triangles nanoantenna to simultaneously analyze the polarization states of the fundamental beam. The effect of the source intensity for TH conversion among the LP and CP states, flipping the nonlinear diffraction orders with incipient new harmonics with intensity dependence and design limitations, has also been addressed. To my knowledge, this is the first approach to upconverted full-Stokes imaging using this type of metasurface design. The benefit will be upconverted polarimetry of MWIR at SWIR, where uncooled efficient detectors are available for high-resolution thermal imaging. This efficient, low aspect ratio, simple design for single step lithographic nanofabrication Ge-metasurface is exempt in TH transmission loss due to the size-effect of nanoantenna for assigning appropriate phase gradient for polarization dependent TH diffraction orders generation, and transmission disparity among the orthogonal polarization states. Integration of this type of metasurface for infrared image upconversion will open new possibilities for identifying intense heat signatures, especially for target recognition in next-generation infrared homing devices for surveillance and missile defense systems.

1. **Introduction:**

Since thermal imaging is passive, based on the target's emissivity, reflectivity and environmental exposure and noise background, nonlinear thermal imaging is not very prominent. However, when the heat source is intense enough, like an explosion site, a missile defense system, and heat exhaust, for those types of heat sources, there might be some possibility to convert that intense signal by a nonlinear optical process to a detectable heat signature [1–5]. Technologies have been continuously developing for the MWIR range for target recognition for interesting applications around 4-micron wavelength, even though it requires image upconversion with complicated phase-matching conditions in the nonlinear medium [6,7]. Earlier work on IR up-conversion for imaging in the 1960s by J. E. Midwinter [8] is still an active area of research [8].

Thanks to the engineered interface of metasurface optics, for the solution of the compact imaging system, even in image upconversion in the same package, overcoming the phase-matching challenge. So the metasurface nanoantenna geometry works out a great deal where nonlinear processes, like high harmonic generation (HHG) [9–11], spontaneous four-wave mixing (SFWM) [12,13], spontaneous parametric down-conversion (SPDC) [14,15] happening efficiently without the fundamental beam, and the upconverted counterpart traveling at the same phases inside the nonlinear medium [12,16,17]. Earlier studies by Tymchenko et.al. of a gradient nonlinear PB metasurface with a thin MQW substrate reported to achieve SHG conversion efficiency up to $2\times10^{-4}$ % [18]. Recent studies by Sarma et. al. have shown a multi-QW heterostructure polaritonic metasurface with very high second-harmonic generation (SHG) efficiency of about $1.5\times10^{-2}$ % for pump intensities of 11 kW/cm$^2$ [19]. However, the complexity of the meta-atom poses a challenge for nanofabrication. Also, some nonlinear conversion processes require precise optical alignment of multiple sources, which poses a challenge for dynamic platforms. So here we search for a simple nonlinear process that generates third harmonics in a single-step lithographic design.

If we look for structural simplicity, a handful of research reports Quasi-BIC resonance [9,10,20–23]. These dielectric optical nanoresonators with low optical loss and absorption yield better control over amplitude and phase for both linear and nonlinear optical processes. The concept of BIC came from Von Neumann, although a century old, has recently been revived in explaining the missing mode from the propagating continuum of light, and their longer lifetime in the resonance yields a high-Q metasurface [9,24–28]. Since the optical nonlinear effect is proportional to light intensity, these trapped electromagnetic modes can enhance the local field to a great extent, which initiates the nonlinear conversion due to high local intensity in the meta-atom structure.

Notable research for the MWIR image upconversion system includes the effect of group velocity mismatch when imaging is done with an ultrashort pulse by Ashik et.al. [29]. Some metasurface researchers are focused on upconverted metalens design from SWIR to the visible range [30–32] out of which Schlickriede et.al. made a great effort to investigate the TH-generated beam focusing. Although thermal imaging can be assumed to be an inefficient technology because of the low intensity from passive and environmental sources, but Metasurface can boost the upconversion efficiency, and with the Stokes imaging, the hidden information can be revealed, which has high potential for removing background noise like heat-seeking missile chases in the open sky with solar background, investigating explosion sites [4,6,33,34]. On the other hand, uncooled, compact, efficient MWIR detectors' shortcomings, especially in terms of dynamic range control and sensitivity, and high cost, still pose challenges for MWIR imaging technology on a compact scale. So, if reasonable power can be upconverted to SWIR, where there are more options for uncooled, compact detector choice, that would be a technological venture [34–36]. Also, considering poor MWIR image resolution, the upconverted imaging can be better in temporal resolution, which requires more research [30,37,38].

Very few works have reported nonlinear conversion with Ge metasurface [28,31], and some with Ge alloy metasurface [39–41]. Germanium is a centrosymmetric material that possesses a diamond cubic crystal structure, which is not useful for second harmonic generation. However, with almost vanishing imaginary part of the refractive index at MWIR range and with the third-order nonlinear susceptibility for $\chi^{(3)} \sim 1.6\times10^{-18}$ m$^2$/V$^2$, it could be an attractive material for MWIR third harmonic generation (THG) [31,41]. So, this paper aims to explore Ge-metasurface's suitability for generating sufficient TH response in an efficient way for image upconversion from MWIR to SWIR. Although there are few wonderful literature for metasurface-assisted upconverted imaging [30,32,42]. To my knowledge, no literature has reported the polarization operation, control, and analysis of the upconverted beam for full Stokes imaging purposes using metasurfaces. In this work, I explored the design metric for polarization decoupling from the TH diffraction orders for upconverted polarization-sensitive imaging, transitioning from MWIR to SWIR. With a brief description of the metasurface design parameter and the nonlinear simulation methodology in Section 2, this article has been structured as follows.

For the nanoantenna design, I started with a triangular metasurface, which type was investigated in my prior work, has the ability for phase control and good transmission in both linear regimes [43] and the nonlinear regime [28]. I employ the design tactics of multimode metasurface with $P > \lambda_{TH}$, and this proves to be better for transmission control and TH conversion efficiency as discussed in Section 3.1.

The height 2100 nm of the Ge metasurface is the minimum and necessary for achieving full phase control for the polarization-sensitive metasurface for both linear and nonlinear regime [44]. For the nonlinear response, the coherent length $L_{coherent}$ (Equation. S17 in the supporting information section) to which the nonlinear efficiency increases and then decreases [41]. So, I checked the height increment effect up to the coherent length in section 3.2 and analyzed the result with the transmission line theory, and then input the result to the coupled wave equation of TH generation in nonlinear optics [45,46].

For the phase gradient metasurface design, the single triangular nanoantenna has been used, which has high efficiency for transmitting the power to the first order when the phase gradient is appropriate [47]. Moreover, with the cross-triangular shape nanoantenna, how the linear polarization states can be exploited efficiently has been discussed in detail in section 3.3 with Stokes parameter calculation.

The figure of merit of the metasurface with the variation of the optical power has been discussed in Section 3.4. To explain the result in more depth, I analyze the fifth harmonic generation (FHG) efficiency and try to draw a conclusion for the diffraction order flip among the TH converted light with the increasing intensity, when conversion to the new harmonic order starts. With the emergence of new harmonic order generation, polarimetry for the high-harmonic beam as a function of the polarization state of the fundamental beam faces the limit of Stokes imaging relying on linear diffraction theory and phase gradient of light and may requires a new theory to explain it.

## 2. Metasurface design parameter and Nonlinear simulation methodology

For the nonlinear simulation of Germanium the refractive index has been taken from fitted material data of Palik ($n_{Ge}$= 4.025 at 4 μm wavelength) and $\chi^{(3)} \sim 1.6\times10^{-18}$ m$^2$/V$^2$ has been used in the Chi3/Chi2 material model in the finite difference time domain (FDTD) simulation package from Ansys-Lumerical Inc. Here $\chi^{(3)}$ for 4 μm wavelength has been checked with the generalized Miller's formula using linear refractive index of different wavelength [41,48,49]. In simulation, we see that the relation of nonlinear power conversion with $\chi^{(3)}$ is linearly proportional, so the slightest change of $\chi^{(3)}$ can be adjusted to the final result as long as the imaginary part of the refractive index of the nonlinear material is negligible. Upon checking the height variation with the metasurface TH conversion efficiency of the different diffraction orders, the final design height is set to 2150 nm. For the lateral dimension of the nanoantenna's, the optimized length for the big triangle L1 = 1478 nm, base B1=568 nm, small triangle L2=820 nm, B2 = 300 nm. The four triangles are positioned to optimize the Maximum TH conversion to the 1st orders of the different incident polarization states, as shown in supporting section S3.1. The position of one triangle crossing another is for creating better resonance and light coupling. For this design, the height and length have better tolerance limits of about 50 nm, but the base parameter must be accurate to within about 6 nm. Besides that, this design is very simple and can be fabricated with single-step lithography, and the gap between all the sides of the nanoantenna remains open. The design periodicity is Px=Py=1600 nm. A 4 μm fundamental wavelength source of 1.17 THz bandwidth at normal incidence is used for all the simulations.

## 3. Simulation Results and Discussion

### 3.1 Nonlinear conversion efficiency for the single-mode and multimode nanoantenna:

For imaging metasurfaces, we prefer a single-mode nanoantenna, so the periodicity is adjusted to be less than the wavelength. For TH conversion, I checked the ratio $\lambda_{TH}/P$ = 0.977 and 1.20 for an equilateral triangle with an incremental ratio of L/P = B/P = 0.375 to 0.937, and the transmitted spectral intensities presented in Figure 1a for both TE (y-polarized) and TM (x-polarized) incident. Looking at the intensities at the 1330 nm wavelength, we see the TH conversion increasing with increasing nanoantenna dimension, but more in magnitude for the $\lambda_{TH}/P$ = 1.20 condition. For efficient TH conversion, I decided to use a periodicity of P = 1600 nm. Figure 1b and 1c provide the quantitative TH conversion efficiency ($\eta_{TH}$ %). From Figure 1c and the supporting information section figure S.1e, we see that for P=1300 nm, the L=B=1218 nm triangle has $\eta_{TH,\ int}$ = $1.89\times10^{-5}$% for TE and $1.92\times10^{-5}$% for TM incidence, respectively, which is the highest for $\lambda_{TH}/P$ = 0.977. For triangle sizes 800, 900, and 1000 nm, the $\eta_{TH,\ int}$ = $3.82\times10^{-6}$ %, $5.17\times10^{-5}$%, and $3.33\times10^{-4}$% respectively for TE incidence. For the same sizes, the $\eta_{TH,\ int}$= $3.02\times10^{-5}$%, $1.09\times10^{-4}$%, and $1.91\times10^{-4}$% respectively for TM incidence. So, for the Ge equilateral triangle of 2050 nm height the $\eta_{TH,\ int}$ is the highest for the triangle size parameter 0.625 for this comparison. Throughout the article, the internal TH conversion efficiency $\eta_{TH}^{int}$ is defined by the ratio of the transmitted intensity at the TH wavelength to the transmitted intensity at the fundamental wavelength:

$$\eta_{TH}^{int} \% = \left(\frac{I_{3\omega}}{I_{\omega}}\right) \times 100 \qquad (1)$$

The $\eta_{TH}^{ext}$ is the same ratio, but instead of transmitted intensity, it is just the incident source intensity at the fundamental wavelength through the finite length $CaF_2$ substrate. As we will see in the next section $\eta_{TH}^{int}$ is our main interest for the quantitative comparison.

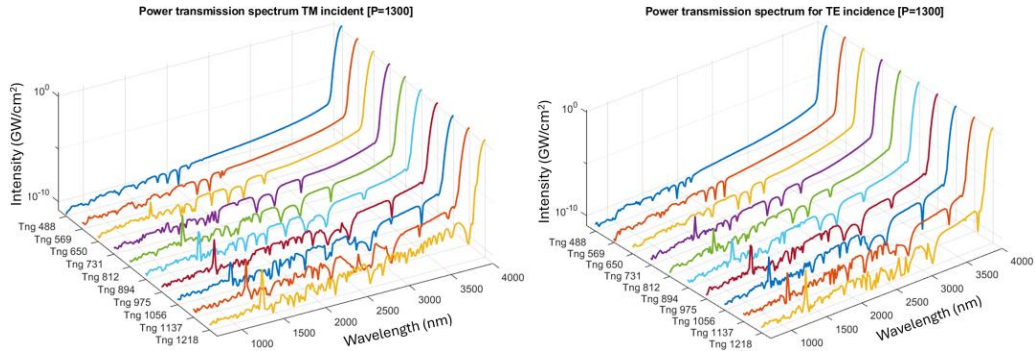

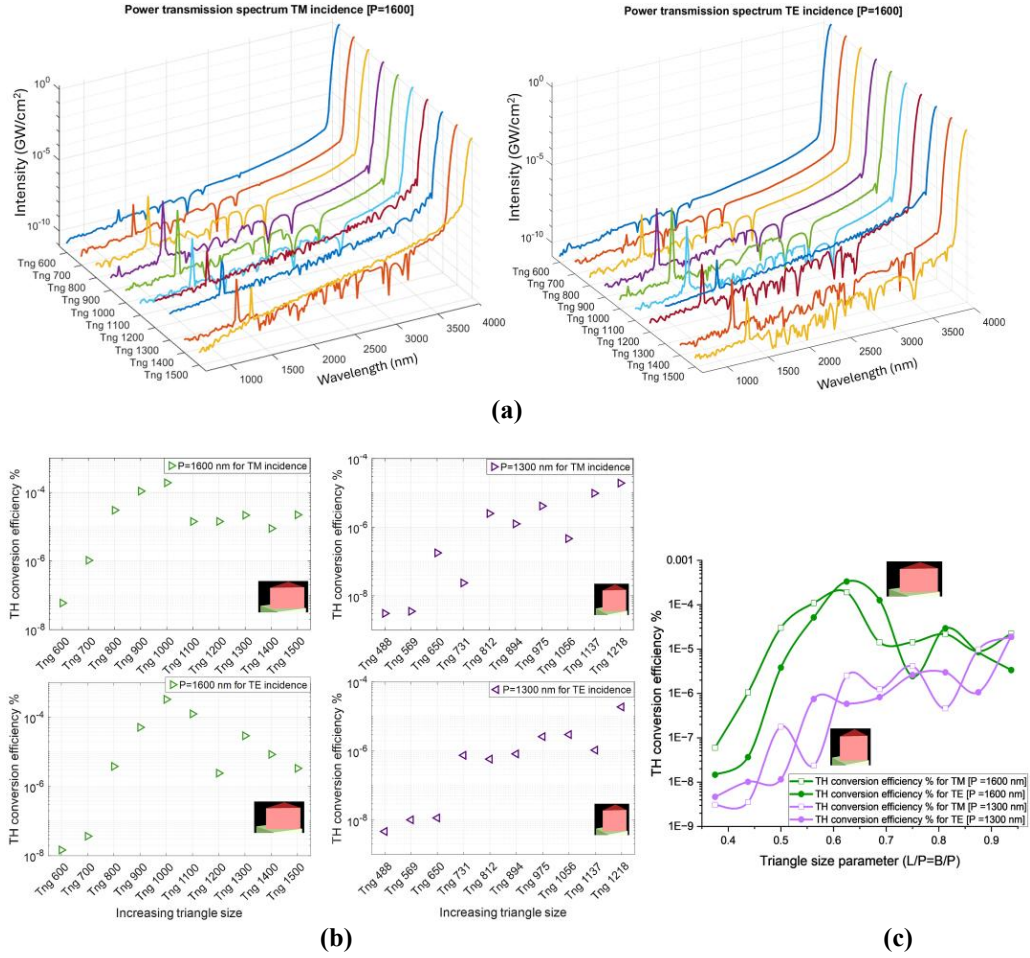

Fig.1. Transmitted intensity comparison between the triangular nanoantenna for two periodicities. (a) Top one for periodicity = 1300 nm (left) TM, (right) TE incidence, respectively. Bottom one for periodicity = 1600 nm (left) TM, (right) TE incidence, respectively. (b) TH conversion efficiency for (left) periodicity = 1600 nm (right) periodicity = 1300 nm for TM incidence. (c) Comparison of the TH conversion efficiency for $\lambda_{TH}/P$ = 0.977 and 1.20 cases both for the TE and TM incidence. These are Ge 2050 nm height equilateral triangles with an incremental ratio of L/P = B/P = 0.375 to 0.937 for both periodicities.

The triangle size parameter 0.625 is associated with the characterized magnetic resonant part of the Mie scattering for the characteristic lateral dimension by L=B ~ $\lambda_F/n_{Ge}$ when the electric field induces the displacement current and strong light coupling occurs [43,50,51]. That might be a reason for the extra $\eta_{TH}^{int}$ for around L=B=1000 nm triangles, apart from that, both the ratio $\lambda_{TH}/P$ = 0.977 and 1.20 cases TE and TM incidences have similarities in $\eta_{TH}^{int}$. Among the other notable features of the size parameter, we see more resonance occur, and the peak is shifted towards the higher wavelength with increasing size parameter. Also, the lack of rotational symmetry for the triangular nanoantenna TE and TM incident response is slightly different even for the single-mode nanoantenna. All this size and shape comparison for TH conversion is strongly shape-dependent [28].

**3.2 The Effect of film height and metasurface height for TH coupling to the higher order:**

To optimize total TH conversion, I examine the effects of Ge film height and metasurface height, as shown in Figure 2a. An unconventional approach of analyzing by the transmission line theory (TL) or the transfer matrix method for nonlinear optical multilayer, I calculated the Ge layer height dependence TH conversion, and compared the result with the FDTD simulated result [52–54]. There are similarities and discrepancies, yet it indicates that a possible assumption can be drawn from TL theory [53,54]. Please see the Supporting Information section S2.1 and Figure S2.1. The coherent length for Ge is calculated to be 2467 nm, up to which TH conversion should rise and then fall. But for a metasurface system, that convention does not hold. There is an indication of the Fabry-Perot enhancement type of behavior, where forward and backward propagating fundamental and TH beams interfere with each other and enhance fundamental beam intensity, and eventually enhance TH conversion at a certain height of the Ge layer [54]. Also, in Figure S2.1, a quasi-phase-matching type of TH intensity behavior has been found with Ge film height increment [16,45].

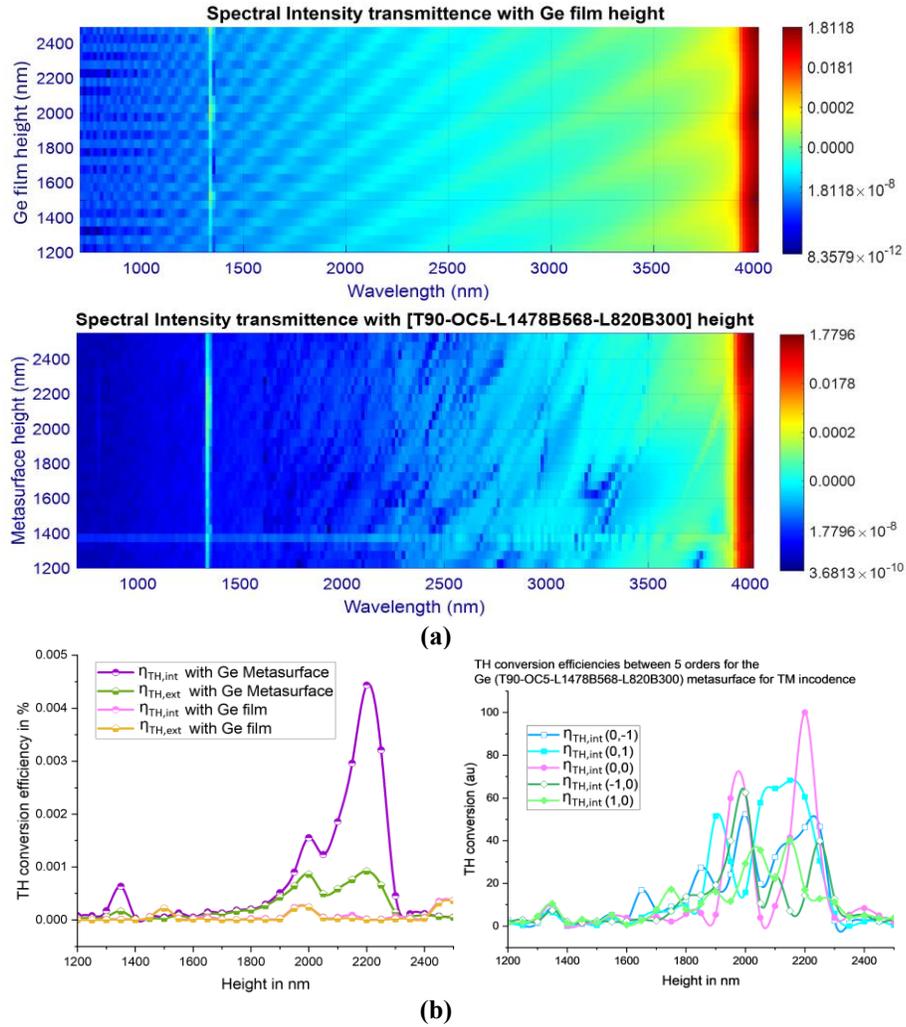

Fig. 2. Spectral intensity transmission with height of (top) Ge film, (bottom) Ge metasurface for LP incidence. The source intensity is fixed 1.87 GW/cm$^2$ at 4μm wavelength. The color bar indicates the transmitted intensity in logarithmic scale value in GW/cm$^2$. (b) The total TH conversion efficiency ($\eta_{TH}$ %) for Ge film and the Ge metasurface with height variation (left).

The diffraction order dependent η$_{TH}$ for this metasurface final design (as shown in Figure 3a, which is named T90-OC5-L1478B568-L820B300, Figure 2a (bottom) and Figure 2b, which shows η$_{TH}$% goes over 4×10$^{-3}$ for 2200 nm height metasurface. However, the diffraction order analysis indicates it is for the (0,0) order, which is not of the interest for pure TH polarization-sensitive imaging. Although it is at a different wavelength and may require a different detector to measure, the (0,0) order position also corresponds to the fundamental beam. So, my design choice is optimized at 2150 nm height where TH intensity to the (0,1) order is maximum.

**3.3 Metasurface design and Stokes parameter for the TH polarization-sensitive imaging:**

The standard norm of creating nanoantenna transmission and phase library for designing the phase gradient metasurface, which I checked with the structural range of 900 nm to 1200 nm length of the lateral dimension, and a good phase control can be obtained. However, the TH converted transmission fluctuates to a very great extent as compared to the linear case. From section 3.1 we see that for the same height, the large lateral volume has more η$_{TH}$ so instead of a small nanoantenna arrangement to create the 2π phase gradient, I used a single nanoantenna to create a 1D phase gradient, which can occupy the maximum volume and yet provide sufficient propagating phase control for diverting power to the 1$^{st}$ order and suppress the orthogonal order [47,55]. Please see the supporting information section S3.1 for details. The immense benefit of using only 1st-order modes, without creating higher orders with periodicity beyond the TH wavelength, is an achievement of the simple design that was a drawback in many literatures [16,22,30,41]. A 90-degree rotation of two antennas crossing at a specific length to achieve equal power magnitudes for orthogonal orders, both for LP and CP incident states, is another achievement that is sometimes difficult to achieve with arrays of smaller nanoantennas. Finally, the two small nanoantennas address the Pancharatnam-Berry (PB) phase requirement for CP light [57,58] deflection with high efficiency (please see section S3.1), which would have required rotating smaller nanoantennas. Under an intense electric field, the bulk material response for the polarizability of the nonlinear medium [45]:

$$P_i = \epsilon_0 \sum_i \chi^{(1)}_{ij} E_j + \epsilon_0 \sum_{jk} \chi^{(2)}_{ijk} E_j E_k + \epsilon_0 \sum_{jkl} \chi^{(3)}_{ijkl} E_j E_k E_l + \cdots \qquad (2)$$

Where E is the local electric field, $\epsilon_0$ is a vacuum permittivity, $\chi^{(1)}$ is a linear susceptibility of the medium, and $\chi^{(2)}$ is the second and $\chi^{(3)}$ is the third-order nonlinear susceptibility tensor, respectively. With increasing field intensity, the nonlinear terms, $\chi^{(3)}$, can become sufficiently large, and the material response becomes nonlinear. For the linear optical regime, $P_i \propto E_j$ and for orthogonal fields $E_x = E_{0x} e^{-i\varphi_x}$ and $E_y = E_{0y} e^{-i\varphi_y}$ with phase difference δ = (φ$_y$ - φ$_x$) the Stokes parameters can be measured as from Equation 3, when we know the intensities [56–59].

$$s = \begin{bmatrix} s_0 \\ s_1 \\ s_2 \\ s_3 \end{bmatrix} = \begin{bmatrix} E_x^2 + E_y^2 \\ E_x^2 - E_y^2 \\ 2E_x E_y \cos\delta \\ 2E_x E_y \sin\delta \end{bmatrix} = \begin{bmatrix} I_x + I_y \\ I_x - I_y \\ I_{L45} - I_{L-45} \\ I_{RCP} - I_{LCP} \end{bmatrix} \qquad (3)$$

For the nonlinear regime, when the THG process originates, $P_i$ is not proportional to the electric field, and the three-photon process requires three incident electric fields $E_j, E_k, E_l$ of frequency ω to generate a TH photon of frequency 3ω. The third term of Equation 2 then [37]:

$$P_i = \epsilon_0 \sum_{jkl} \chi^{(3)}_{ijkl} E_j E_k E_l = \chi^{(3)}_{out\, in} \psi^{(3)}_{in} \qquad (4)$$

The j, k and l indices are the polarization direction of the incoming light, which, for the case of a purely polarized incoming fundamental frequency beam, can be written as "in" and the index 'i' is for the outgoing TH beam 'out'. For the coherent incident fundamental frequency beam, the electric field state vector's polarization state can be written as [37,60]:

$$\psi^{(3)}_{in}(\omega, \omega, \omega) = \begin{bmatrix} E_1^3 \\ E_2^3 \\ 3 E_1^2 E_2 \\ 3 E_1 E_2^2 \end{bmatrix} \qquad (5)$$

The triple Stokes vector can be written as:

$$\psi^{(3)}_N(3\omega) = Tr(\rho^{(3)} \gamma_N) \qquad (6)$$

Where $\rho$ is the 4×4 coherency matrix and $\gamma$ is the analogue of the Pauli matrices. From this, the 16×1 triple Stokes vector that presents the incident fundamental beam in terms of the four Stokes components can be written as:

$$S^{(3)}_N(\omega, \omega, \omega) = \begin{bmatrix} \sqrt{2}s_0(5s_0^2 - 3s_1^2) \\ \sqrt{6}(-\frac{4}{3}s_0^3 + 3s_0^2 s_1 + 2s_0 s_1^2 - 3s_1^3) \\ \sqrt{3}(-\frac{8}{3}s_0^3 - 3s_0^2 s_1 + 4s_0 s_1^2 + 3s_1^3) \\ s_1(3s_0^2 + s_1^2) \\ s_2(s_2^2 - 3s_3^2) \\ 3(s_0 - s_1)(s_2^2 - s_3^2) \\ 9s_2(s_2^2 + s_3^2) \\ 3s_2(s_0 - s_1)^2 \\ 3s_2(s_0 + s_1)^2 \\ 3(s_0 + s_1)(s_2^2 - s_3^2) \\ s_3(3s_2^2 - s_3^2) \\ -6s_2 s_3(s_0 - s_1) \\ 9s_3(s_2^2 + s_3^2) \\ -3s_3(s_0 - s_1)^2 \\ 3s_3(s_0 + s_1)^2 \\ 6s_2 s_3(s_0 + s_1) \end{bmatrix} \qquad (7)$$

For a PIPO (Polarization-in Polarization-out) measurement system, the polarization states of the light incident through the incident polarizer can be represented by this 16×1 triple Stokes vector. After the light passes through the nonlinear medium and is collected by the analyzer, it can be given by the 4×1 Stokes vector. So, there will be sixteen coordinates with (Ω, ψ)=(latitude, longitude) on the Poincaré sphere for the fundamental beam. Finally, the TH Stokes vector can be presented by these coordinates as [37,60–62]:

$$s = \begin{bmatrix} 1 \\ \cos(2\psi)\cos(2\Omega) \\ \sin(2\psi)\cos(2\Omega) \\ \sin(2\Omega) \end{bmatrix} \qquad (8)$$

Earlier research on TH imaging of biological samples has employed this type of Stokes reconstruction to recover images for isotropic and anisotropic nonlinear media using Stokes-Muller polarimetry. Another interesting quantum mechanical approach to come to the same formalism has been taken by Shaji et.al., where the Stokes vectors are obtained by the operation of the Stokes operator on the photon density matrix, which is composed of annihilation $a_H^{\pm}$ and creation operator $a_V^{\pm}$ [62]. This three-photon formalism by Samim et.al [37], Shaji et.al [62], Sar et.al. [61], Kontenis et.al. [60] serves for third harmonic generated high resolution imaging for biological samples using Stokes-Muller polarimetry [63,64]. After that, metasurface optics has revolutionized the imaging system to the compact scale with multifunctionality in every aspect [56–58,65–68], in which the intensities at the diffracted order position provide the six or at least four of these intensity magnitudes, depending on the incident polarization states of the light.

Yet Stokes parameter reconstruction of upconverted image or using diffraction orders for nonlinear regime polarimetry has not been in any research focus or discussion in the literature to this date, except for very few [69], but not with metasurface. The metasurface of the nonlinear material provides both intricacy and efficacy because birefringence is not a material property, but rather a structural accessory that can be engineered. The phase gradient maneuvering wavevector control, which bypasses the stringent phase matching condition and different resonant conditions, enables much more control over various nonlinear processes.

Figure 3(a) represents the high-efficient phase gradient nonlinear Ge-metasurface design unit cell, which functions as an image upconverter and four-polarization-state analyzer simultaneously by sending the TH beam to the prominent orders at $\lambda_{TH}$ = 1330 nm. In Figure 3c, the detectors pixel's angular position for TM at D1(0,-1), TE at D2(1,0), LCP at D3(1,0), RCP at D4(-1,0), and all (0,0) order positions indicated sufficient spatial separation in the detector plane for collecting optical power with FPA at TH wavelength. The $s_{TH_0}^{3\omega}$ is the total intensity, $s_{TH_1}^{3\omega}$ is the difference in the intensity component $I_{x,TH}^{3\omega}$ and $I_{y,TH}^{3\omega}$, $s_{TH_3}^{3\omega}$ is the difference in the intensity of the component $I_{RCP,TH}^{3\omega}$ and $I_{LCP,TH}^{3\omega}$. Since this metasurface has only five output orders, this design can measure $s_{TH_1}^{3\omega}$ effectively by blocking L45 light completely for one orthogonal basis. For calculating the TH Stokes parameter (Figure 3b and Figure S4.1), $s_{TH_1}^{3\omega}$ and $s_{TH_3}^{3\omega}$ have been weighted by the total intensity from the four order positions for each incident polarization state for scaling the Stokes parameter from 0 to 1. This is a solution for experimental conditions for reconstructing $s_{TH}^{3\omega}$, with proper calibration for the metasurface instrumental matrix, since no theory is available yet to predict the nonlinear interaction for the electric fields and the diffraction theory for a strongly resonant metasurface, for that matter. From the $s_{TH}^{3\omega}$ values at each detector pixel, the degree of polarization (DOP) can be calculated:

$$DOP = \frac{(s_{TH_1}^{3\omega}{}^2 + s_{TH_2}^{3\omega}{}^2 + s_{TH_3}^{3\omega}{}^2)^{\frac{1}{2}}}{s_{TH_0}^{3\omega}{}^2} \qquad (9)$$

Here, Ge nanoantenna has been simulated with an isotropic susceptibility, but the structural birefringence is an issue for the metasurface, which may need the effective $\chi_{ijkl}^{(3)}$ modeling from nonlinear diffraction theory [70] and nonlinear Raman–Nath diffraction modeling [71,72]. For modeling the triple Muller matrix M[(3)], all the components of the outgoing Stokes vector $s_{TH}^{3\omega}$

at all the sixteen coordinate points of the three-photon Poincaré sphere need to be measured [37]:

$$M^{(3)} = s^{3\omega}_{TH} \ s^{-1} \quad (10)$$

Where $s^{3\omega}_{TH}$ is a 4×16 matrix, and $s^{-1}$ is the 16×16 matrix. For the parametric upconversion with sum frequency generation (SFG) polarization imaging with lithium niobate film a Stokes-Mueller formulism has been carried out by Zhu et. al. where 4×4 Mueller matrix is the function of both the nonlinear susceptibility and pump polarization [73]. A calibration by the measured SF Poincaré sphere can retrieve the FF polarization state. For the metasurface design of this article the reconstructed $s^{3\omega}_{TH}$ for the four maximally different input pure polarization states of FF has been shown in Figure 3d to compare the one-to-one mapping of FF and TH Stokes vectors.

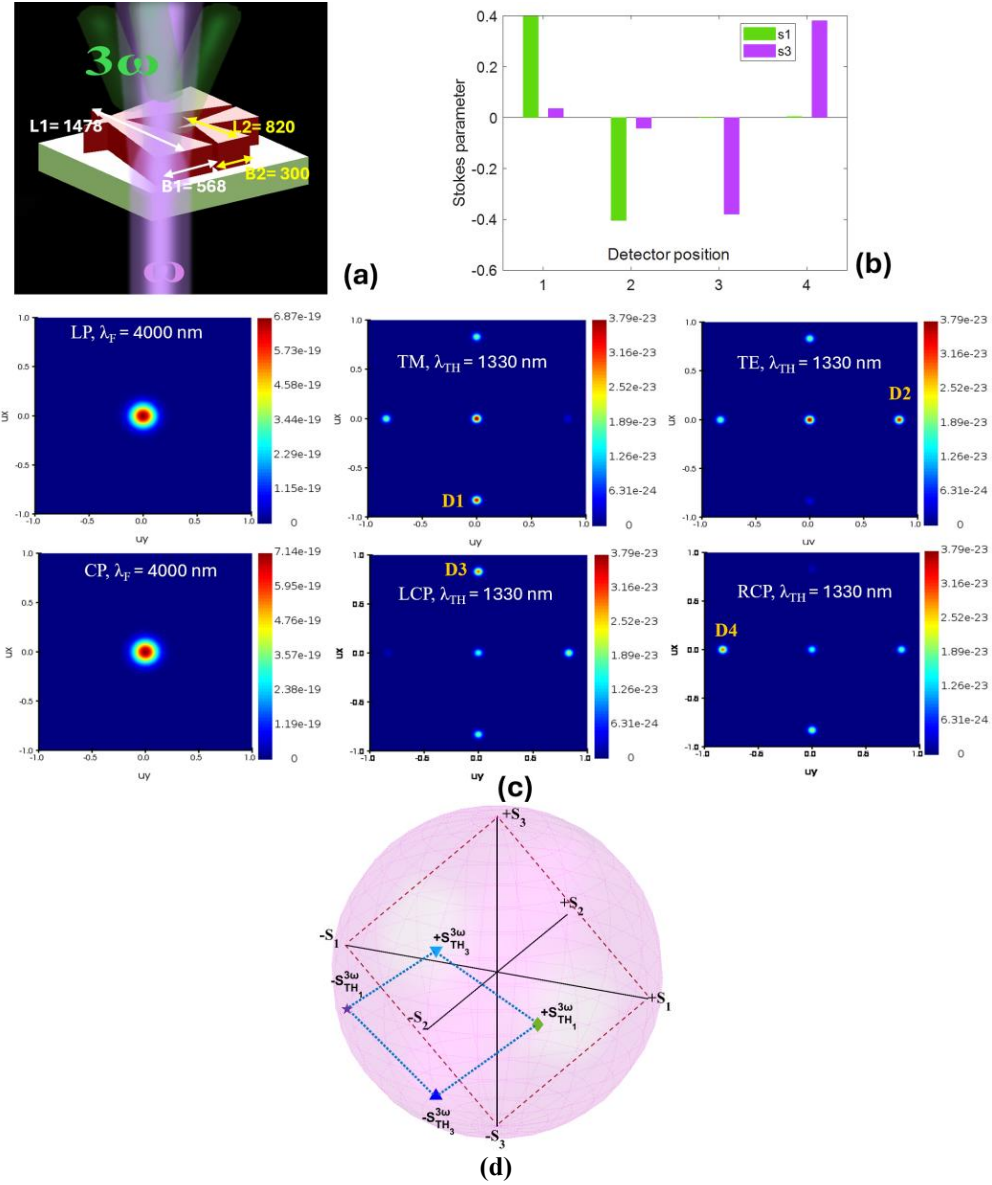

Fig. 3. (a) Metasurface unit cell design [name T90-OC5-L1478B568-L820B300] with Length (L) and Base (B) values for big triangle and small triangles. (b) Calculated Stokes parameter values for the incident intensity of 1.87 GW/cm². (c) Optical power transmission in the far-field for λ_FF and λ_TH. The color bar indicates the transmitted optical power. The four prominent orders at λ_TH = 1330 nm, for TM at D1(0,-1), TE at D2(1,0), LCP at D3(1,0), RCP at D4(-1,0) positions, and all (0,0) order position. The Intensity from these prominent order positions are used to calculate the $s_{TH}^{3\omega}$ Stokes parameter as shown in (b). (d) The FF Stokes vector (connected by the dashed red line) and the corresponding $s_{TH}^{3\omega}$ points in the Poincaré sphere for the metasurface-analyzed TH intensities (connected by a blue dotted line) from all the order positions.

The TH diffraction angle is determined by [32]:

$$sin(\theta) = \frac{k_{x,y}}{k_{TH}} = \frac{\lambda_{TH}}{P_{PG}(x,y)} \qquad (11)$$

Where, $k_{TH} = \frac{2\pi}{\lambda_{TH}}$ and $k_{x,y} = \frac{2\pi}{P_{PG}(x,y)}$, $P_{PG}(x,y)$ is the local phase period. Figure S3.2 depicts the beam deflection angles for the first order: 56.24°, 43.01°, and 35.33° for periodicities of 1600, 1950, and 2300 nm, respectively. The TH converted beam projected to the different pixel positions of the detector, shown in Figure 3(b) here. For fundamental TM, TE, LCP and RCP polarized beam the detector positions D1(θ,ϕ)=(56.24°, 180°), D2(θ,ϕ)=(56.24°, 90°), D3(θ,ϕ)=(56.24°, 0°), D4(θ,ϕ)=(56.24°, -90°), are the TH beam's prominent orders respectively from which the intensity component of $I_{x,TH}^{3\omega}$, $I_{y,TH}^{3\omega}$, $I_{LCP,TH}^{3\omega}$ and $I_{RCP,TH}^{3\omega}$ will be recorded for the TH Stokes parameter analysis at λ_TH = 1330 nm wavelength.

### 3.4 Incident intensity-dependent variation of TH and FH generation and fluctuation of Stokes parameter:

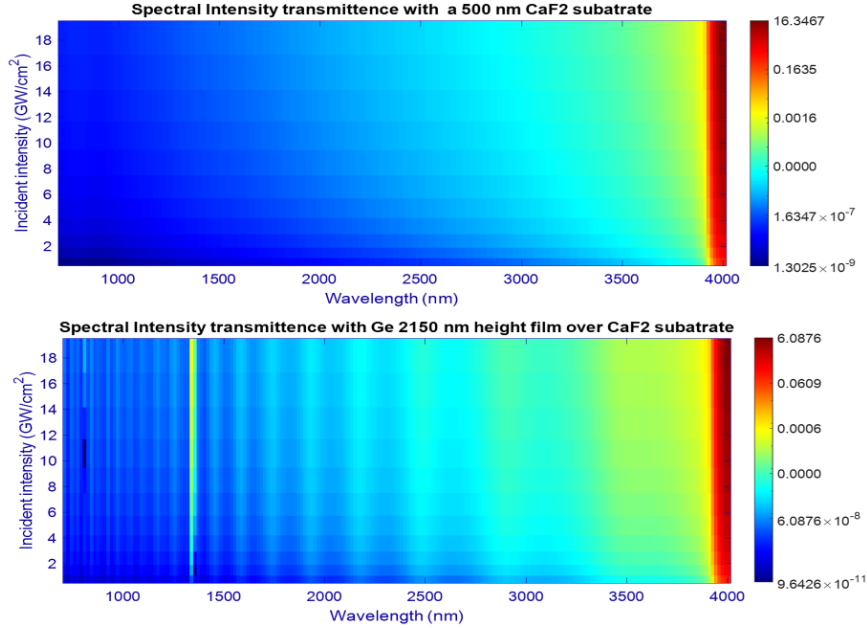

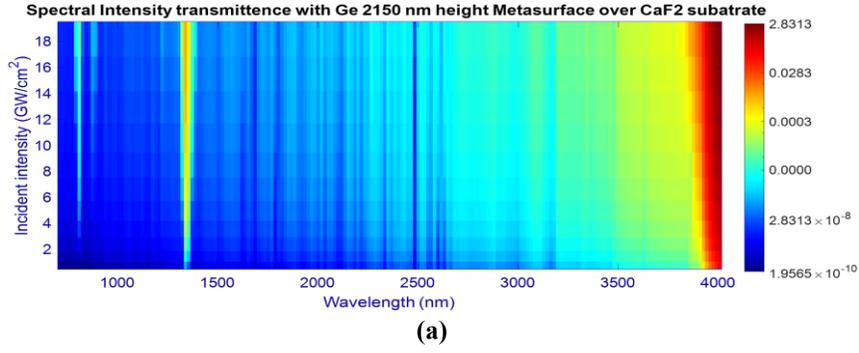

(a)

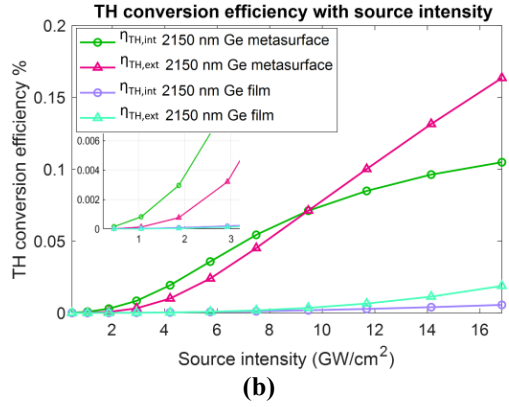

(b)

Fig. 4. (a) Spectral intensity transmission with source intensity (SI), (top) only CaF$_2$ substrate, (middle) Ge film (2150 nm height) over that substrate, (bottom) Ge metasurface (2150 nm height). All for LP incidence. The source intensity ranges from 0.47 GW/cm² to 16.8 GW/cm² at 4μm wavelength. The color bar indicates the transmitted intensity on a logarithmic scale in GW/cm². (b) The total TH conversion efficiency ($\eta_{TH}$%) for the Ge film and the Ge metasurface as a function of source intensity. The inset magnifies the region of low incident power.

The local field amplitude, which can be significantly affected by increasing incident power, overlapping resonance, and interference of the forward and backward fields in the nonlinear medium, is governed by the equation that governs the material's polarization under the influence of higher-order field vectors. Even in the simulation, the $\chi^{(3)}$ for Ge is isotropic, but the field vectors $E_j E_k E_l$ contributed to the different magnitude by the metasurface asymmetric nanoantenna, because of uneven enhancement of the local field. In Figure 4a, the spectral response with increasing source intensity for the finite CaF$_2$ substrate, Ge 2150 nm height film over that substrate, and the same height Ge metasurface over that substrate is depicted, where the transmitted intensity can be seen in logarithmic scale for all cases from the color bar. All are for TM incidence, so we observe an enhanced nonlinear behavior of the metasurface with the third and fifth harmonics (FH starts from 2.92 GW/cm²) in response to a linearly polarized fundamental beam. The internal TH conversion efficiency tends to saturate earlier with increasing source intensity, as depicted in Figure 4b.

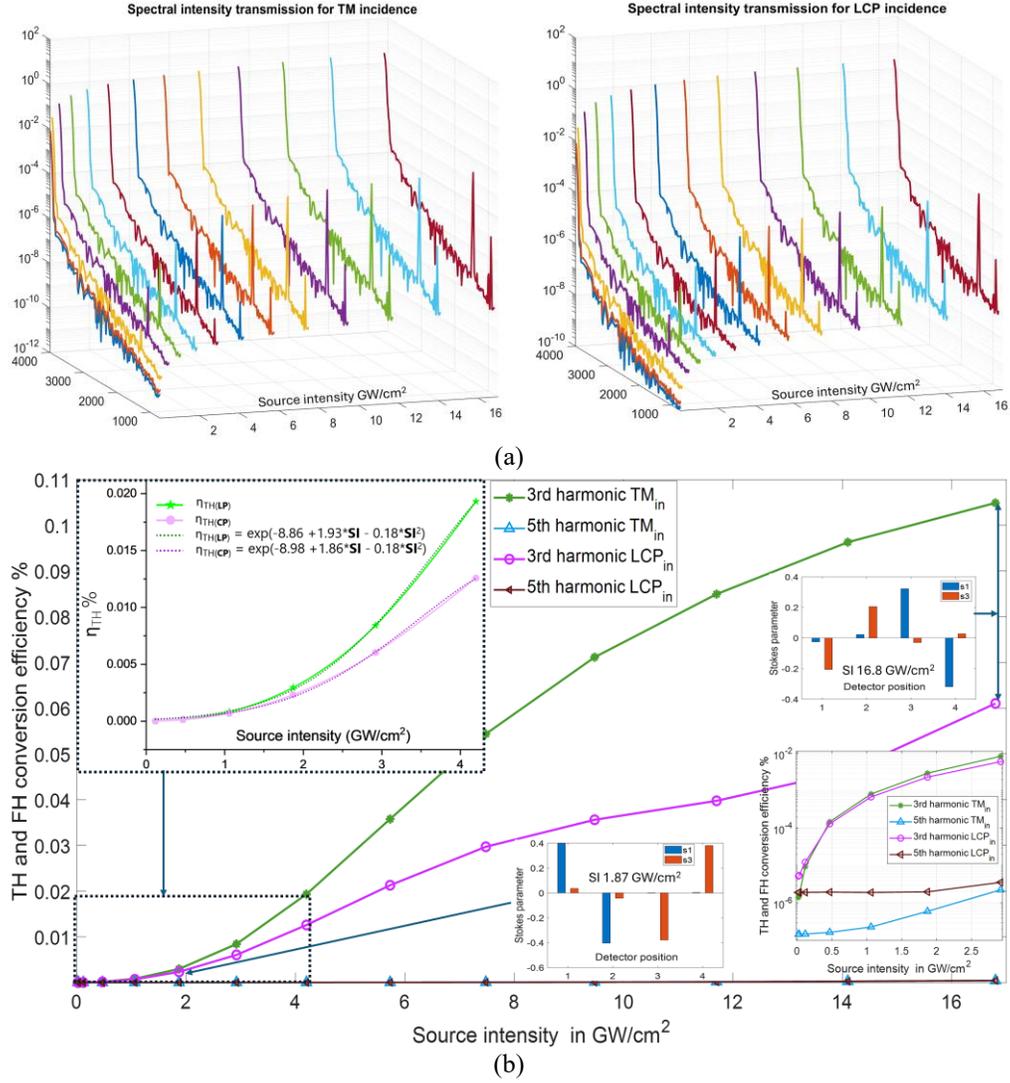

Fig. 5. (a) Spectral intensity transmission through the Ge-metasurface with increasing source intensity. Here, fourteen source intensity levels are increasing from left to right. The peak at 1330 nm of wavelength corresponds to 3rd harmonic generation (THG), and the peak at 799.5 nm of wavelength (which becomes more prominent with increasing incident intensity) corresponds to 5$^{th}$ harmonic generation (FHG). Left side for LP states and right side for CP states. (b) The TH and FH conversion efficiencies $\eta_{TH}$ for LP (green curve) and CP (magenta) states. The inset in the top left corner is the curve fitting of the $\eta_{TH}$(SI), of source intensity (SI) shown in the dotted line for matching exponent of the quadratic polynomial [y = exp(a+bx+cx$^2$)] behavior of the lower intensity range (0.117 - 4.2 GW/cm$^2$) of the source. The right bottom corner shows in a logarithmic scale all the TH and FH efficiencies, so both $\eta_{TH}$ and $\eta_{FH}$ are visible for the source intensity range (0.117 - 2.92 GW/cm$^2$) where the Stokes imaging works in the desired detector position as indicated in Figure S4.1 and S4.2(b) in the supporting information section. The bottom and top Stokes parameters correspond to source intensities of 1.87 GW/cm² and 16.8 GW/cm², respectively.

In Figure 5a, I plotted the spectral intensity response of the metasurface for fourteen source powers, considering both LP and CP incidence. From this, the TH and FH conversion efficiencies for both LP (green line) and CP (pink line) are compared as a function of source intensity (SI). If we look at the source intensity range (0.117 - 2.92 GW/cm$^2$), the behavior of both LP and CP are the same. This is the range where the metasurface performs for Stokes imaging, as presented by the reconstructed Stokes parameter $s_{TH}^{3\omega}$ from the simulation. Please see the Supporting Information section S4.1, which highlights the optical intensity level. This range of the TH conversion efficiency has been fitted for understanding the nonlinear behavior, which is represented in the inset at the top left corner. The $\eta_{TH}$ comes as an exponential raising function of a quadratic polynomial of SI [y = exp(a+bx+cx$^2$)], as indicated with dotted green and magenta lines. The coefficients are a ≈ -8.9, b ≈ 1.9, c ≈ -0.18, and although the c value is low, for both LP and CP cases, the curve deviated from a simple exponential rise. One conclusion is that the metasurface-generated TH efficiencies do not follow a simple power dependence. Please see the supporting information section S4.2 for details of the logarithmic fit, so the [log ($\eta_{TH}$) ∝ -8.9+1.9 SI+(-0.18) SI$^2$] as shown in the inset of the bottom right corner. Examining the supporting information section S4.1, one question remains: why does the Stokes parameter change the diffraction order position between the LP and CP states with increasing optical power? This metasurface has been demonstrated to be highly efficient compared to Ge film by 1.4×10$^{-4}$ % to 9.93×10$^{-2}$ % for LP states and 1.2×10$^{-4}$ % to 5.54×10$^{-2}$ % for CP states at incident intensity levels of 0.468 GW/cm$^2$ and 16.83 GW/cm$^2$, respectively, as we can see from Figure 4b and Figure 5b.

However, this limits the intensity range of the Stokes-imaging operation, which will pose an extra burden for the detector's dynamic range and input adjustment for real-world TH Stokes passive thermal imaging. Let's examine the very low intensities below the yellow-highlighted area at 0.0187 GW/cm² and 0.0187 GW/cm² in Figure S4.2 of the supporting information. We observe that the LP and CP TH conversions deviate significantly, and that low intensity $\eta_{TH, CP}$ is associated with a higher $\eta_{TH, LP}$, because the fifth harmonic (FH) generation is more than an order of magnitude lower for CP states compared to LP (as shown by the blue and dark brown curves). Also, this is evident in Figure 5a. Additionally, within this range of intensities, the Stokes parameter reverses the designated order position. CP goes to the LP order position (Figure S4.1). The opposite happens above 2.92 GW/cm$^2$ when LP goes to CP's designated order position, when FH starts to rise almost exponentially. Therefore, we can conclude that when nonlinear orders emerge and undergo changes in the dependence of the incidence power, which affects the diffraction order selection, linear diffraction theory may not be sufficient. So, the concluding takeaway from this discussion is that although harmonic generation is the nonlinear material's property, we have to be more adept in nonlinear metasurface design to suppress undesirable and enhance desirable harmonics for upconverted Stokes imaging.

## 4. Conclusion

The feasibility of metasurface-assisted third-harmonic full-Stokes imaging has been explored with a Ge cross-triangular nanoantenna. The nonlinear conversion enhancement, which tunes the size parameter of the nanoantenna, and the comparison of all nonlinear efficiencies with the Ge film, show a promising route for thermal image upconversion with the Ge metasurface. Along the way, the TH conversion efficiency has been modeled for Ge film thickness variation as an isotropic nonlinear medium, using transmission line theory as input to the coupled wave equation for TH generation. The result has been compared with the FDTD simulation. For imaging metasurfaces, the multimode nanoantenna introduces complexity. In an upconverted imaging scenario, creating a phase gradient requires periodicity higher than the harmonic wavelength, making the unavoidable multimode a design problem. Here, I leverage this into a design advantage, where upconversion is accompanied by polarization decoupling in higher orders, achieved through carefully designed nanoantenna geometry. Therefore, I reconstructed the TH upconverted Stokes parameter and demonstrated that both linear and circular polarization states of the fundamental beam are recoverable over a range of source intensities. Additionally, the limitation of this operation beyond the range of optical power has been addressed, which could be a new direction of research in terms of nonlinear diffraction theory in terms of phase gradient and controlling higher harmonic generation and suppression through resonance-enhanced nanoantenna geometry, as controlling nonlinear material properties is more challenging. Finally, although this highly efficient metasurface has its limitations in operation, with proper calibration and the ability to tune incident power in a calibrated environmental background, it can be a new option for upconverted full-Stokes thermal imaging from MWIR to SWIR, enabling infrared homing and target recognition for intelligent missile defense systems and long-distance surveillance.

## 5. Back matter

**Funding:** This research received no external funding.

**Disclosures.** The author declares no competing financial interest.

**Data availability:** No data was generated or analyzed in the research presented.

**Supplemental document.** Please see the associated "Supporting document for Third Harmonic Upconverted Full-Stokes Imaging with High-Efficiency Germanium Metasurface from MWIR to SWIR."

# Supporting Information: Third Harmonic Upconverted Full-Stokes Imaging with High-Efficiency Germanium Metasurface from MWIR to SWIR.

**HOSNA SULTANA**

*School of Electrical & Computer Engineering, University of Oklahoma, 110 W Boyd St, Norman, OK 73019, USA*
*hosna@ou.edu*

## 1.1 TH CONVERSION EFFICIENCY WITH THE LENGTH SCALE OF THE TRIANGULAR NANOANTENNA FOR TWO PERIODICITIES.

To compare the TH conversion efficiency for the generalized length scale (L/P = B/P = 0.923) of an equilateral triangular nanoantenna with a height of 2050 nm, the spectral intensity transmission has been checked using Ansys-Lumerical Inc. FDTD for periodicities of P = 1300 nm and P = 1600 nm.

Since P < $\lambda_{TH}$ has a single transmission order, P > $\lambda_{TH}$ has a minimum of five transmission orders for TH to split. Therefore, we first attempt to determine the total TH magnitude for both TE and TM incident cases in Figure S1.

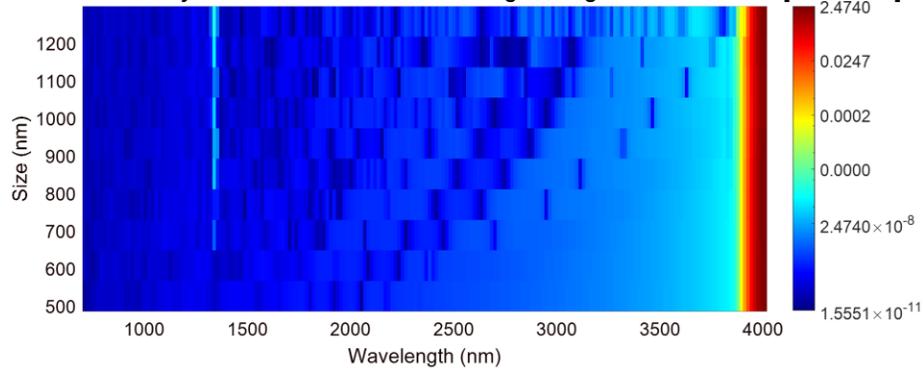

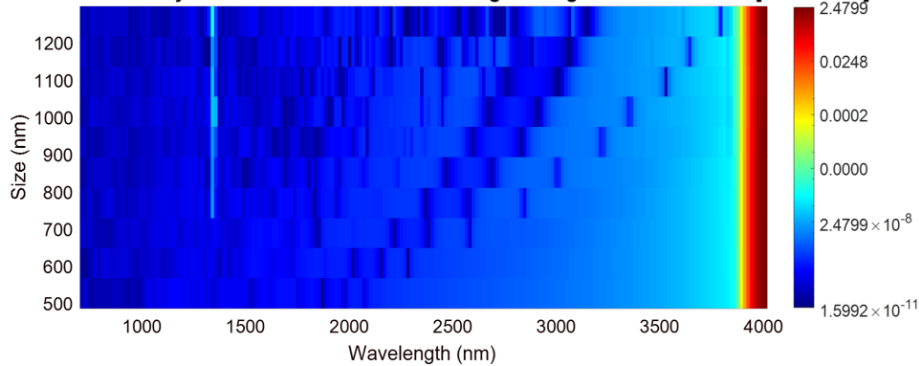

(a)

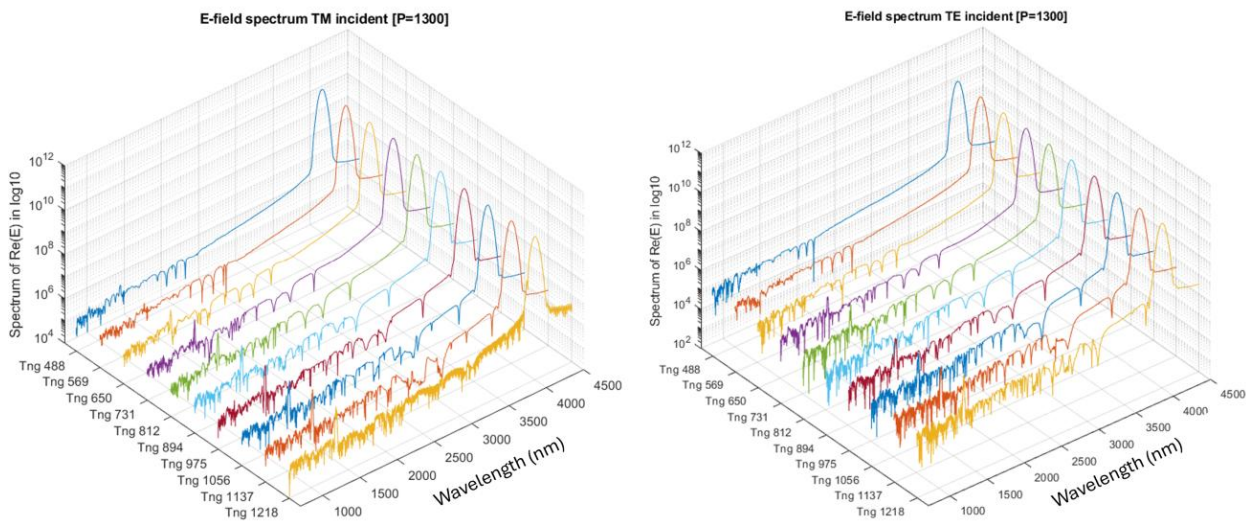

(b)

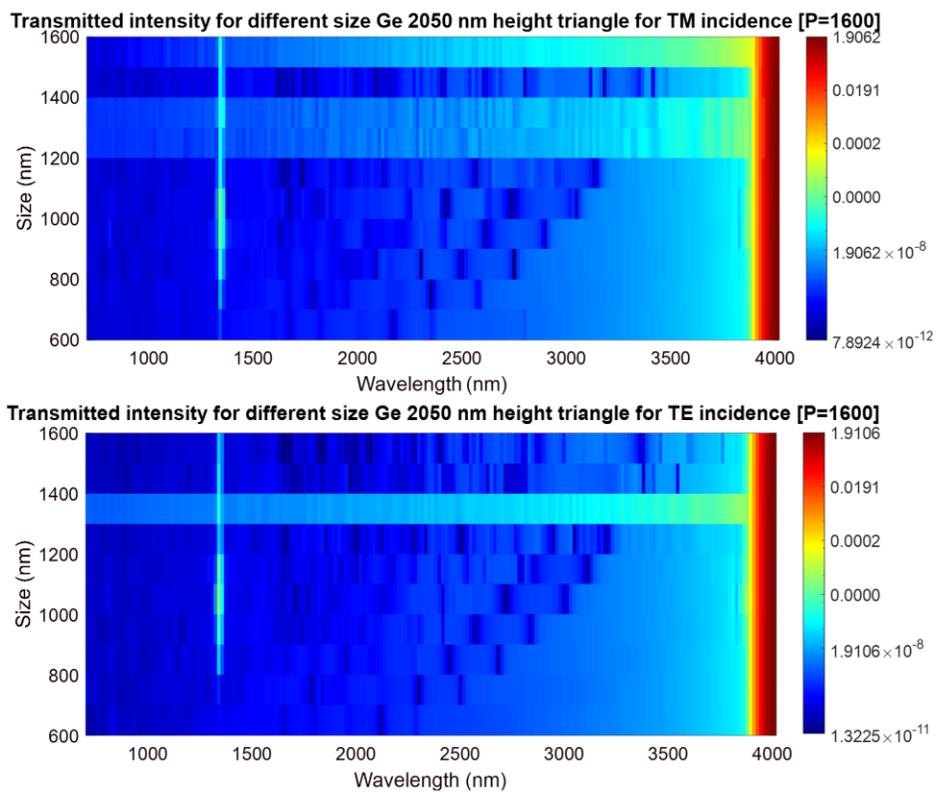

(c)

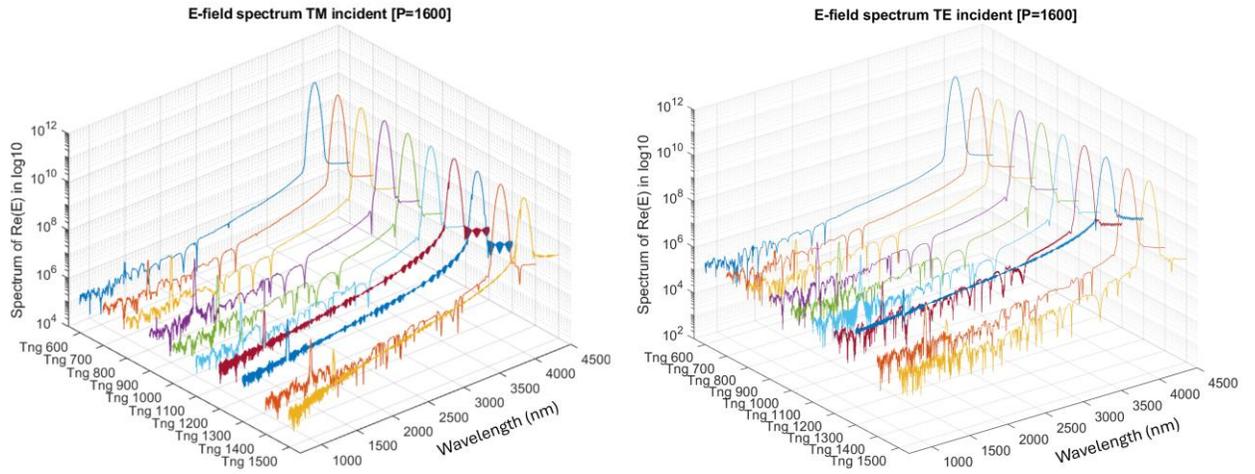

(d)

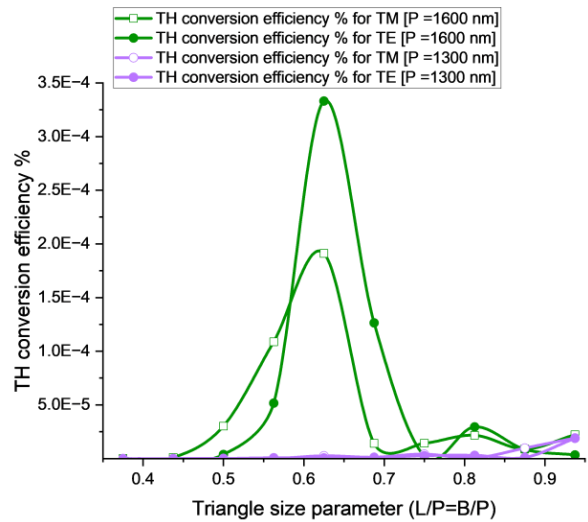

(e)

Fig S1.1. (a) The spectral intensity transmission for different sizes parameter [L=B= 488 nm to 1218 nm] nanoantenna for P = 1300 nm for both TM (top) and TE (bottom) incidence of the fundamental beam of 4000 nm wavelength. The color bar indicates the transmitted intensity in logarithmic scale value in $GW/cm^2$. (b) Real part of the transmitted E-field in log scale for both TM and TE incident for (a). (c) The spectral intensity transmission for different sizes [L=B= 600 nm to 1500 nm] nanoantenna for P = 1600 nm for both TM (top) and TE (bottom) incidence of the fundamental beam of 4000 nm wavelength. The color bar indicates the transmitted intensity in logarithmic scale value in $GW/cm^2$. (c) Real part of the transmitted E-field in log scale for both TM and TE incident for (c). (e) Triangular nanoantenna's size-dependent TH conversion efficiency.

## 2.1. Modeling a multilayer system with transmission line (TL) theory for a nonlinear optical medium:

The equivalent circuit method models electromagnetic scattering by a multilayer and the 2D antenna surface with the transmission line (TL) theory and the lumped circuit parameter of electrical energy propagation. Like lumped circuit elements for analyzing electrical components, there might be some approaches to modal optical gratings, metasurfaces, and photonics integrated circuits, which are still limited due to a lack of accuracy compared to other numerical models of full-wave simulation, like FDTD and FEM. Considering the homogeneous dielectric layers system, which has no structural discontinuities in the lateral direction, the system can be modeled using the transmission line (TL) theory with the transfer matrices of the consecutive layers [1–3].

$$\begin{bmatrix} E_{m+1}^+ \\ E_{m+1}^- \end{bmatrix} = \begin{bmatrix} T_{11} & T_{12} \\ T_{21} & T_{22} \end{bmatrix} \begin{bmatrix} E_m^+ \\ E_m^- \end{bmatrix} \quad S(1)$$

This T-matrix, also known as the ABCD matrix, is compiled by multiplying the N consecutive T matrices for N homogeneous dielectric layers, yielding the output amplitude from the input.

$$\begin{bmatrix} T_{11} & T_{12} \\ T_{21} & T_{22} \end{bmatrix} = [M_1] \dots [M_N] \quad S(2)$$

$$[M_n] = \begin{bmatrix} \cos(\theta_n) & i\sin(\theta_n) Z_n \\ \dfrac{i\sin(\theta_n)}{Z_n} & \cos(\theta_n) \end{bmatrix} \quad S(3)$$

These matrices ensure the reciprocity of the system. The input impedance and the transmission or reflection coefficients can be retrieved as:

For a single layer of Ge over the substrate:

$$\begin{bmatrix} T_{11} & T_{12} \\ T_{21} & T_{22} \end{bmatrix} = \begin{bmatrix} \cos(\theta_{subs}) & i\sin(\theta_{subs}) Z_{subs} \\ \dfrac{i\sin(\theta_{subs})}{Z_{subs}} & \cos(\theta_{subs}) \end{bmatrix} \begin{bmatrix} \cos(\theta_{Ge}) & i\sin(\theta_{Ge}) Z_{Ge} \\ \dfrac{i\sin(\theta_{Ge})}{Z_{Ge}} & \cos(\theta_{Ge}) \end{bmatrix} \quad S(4)$$

Here, $Z_{subs} = \dfrac{Z_0}{\sqrt{\epsilon_{subs}}}$, $\theta_{subs} = \dfrac{2\pi}{\lambda_0} \sqrt{\epsilon_{subs}} H_{subs}$, $Z_{Ge} = \dfrac{Z_0}{\sqrt{\epsilon_{Ge}}}$, $\theta_{Ge} = \dfrac{2\pi}{\lambda_0} \sqrt{\epsilon_{Ge}} H_{Ge}$

Where $\sqrt{\epsilon_{Ge}} = n_{Ge}$ and $H_{Ge}$ is the complex permittivity and height of the Ge layer, respectively, and $Z_0$ is the free space impedance. As shown in Figure 1, the calculated results of the TL method (left) and the FDTD simulated result (right) for a finite CaF2 substrate and an increasing height of the Nonlinear Ge film exhibit some common trends.

For a nonlinear medium response [4,5]:

$$n_{Ge} = (n_{o,Ge} + \Delta n_{2,Ge} |E|^2) - j(k_{o,Ge} \pm \Delta k_{2,Ge} |E|^2) \quad S(5)$$

Where, $\Delta n_{2,Ge} = \dfrac{1}{2} n_{2,Ge} c\, n_{o,Ge} \epsilon_0 |E|^2$

and the third-order nonlinear coefficients

$$n_{2,Ge} = 3\frac{\chi^{(3)}}{4\,n_{0,Ge}^2\,\epsilon_0\,c} \qquad S(6)$$

Here, $n_{0,Ge}$, and $k_{0,Ge}$ are the frequency-dependent linear real and imaginary parts of the refractive index, $\Delta n_{2,Ge}$ and $\Delta k_{2,Ge}$ is the nonlinear part of the refractive index with a Kerr-type nonlinearity. The third-order nonlinear susceptibility is $\chi^3$, and $\epsilon_0$ is the vacuum permittivity, c is the speed of light, and E is the Electric field of the incident light. Putting Eq. S(5) and S(6) with zero imaginary part for Ge at the fundamental wavelength, we can compute the components of the T matrix and calculate the input impedance and transmission and reflection coefficient for these CaF$_2$ and Ge layers.

$$Z_{in} = \frac{T_{11}Z_0 + T_{12}}{T_{21}Z_0 + T_{22}} \qquad S(7)$$

$$t = \frac{2Z_0}{T_{11}Z_0 + T_{12} + T_{21}Z_0^2 + T_{22}Z_0} \qquad S(8)$$

$$r = \frac{T_{11}Z_0 + T_{12} - T_{21}Z_0^2 - T_{22}Z_0}{T_{11}Z_0 + T_{12} + T_{21}Z_0^2 + T_{22}Z_0} \qquad S(9)$$

The optical transmission T, reflection R, and absorption A can be found by:

$$T = |t|^2, \quad R = |r|^2, \quad A = 1 - T - R$$

In Figure S2.1(a), the upper left corner shows the plot of Ge height-dependent T, R, and A for this CaF2 substrate and Ge film, as predicted by this TL theory. For comparison, we also present the FDTD-calculated T from the nonlinear simulation. All these plots are for a constant source intensity of 1.87 GW/cm$^2$. Therefore, we can conclude that there is a more substantial fluctuation in the fundamental wavelength within the Ge layer with varying height, similar to the Fabry-Pérot effect, which indeed causes the TH efficiency fluctuation that will be calculated as follows.

Consider this as the plane wave incident on the substrate and then passing through the nonlinear Ge film:

$$E(z,t) = A_\omega\,e^{i(kz-\omega t)} + c.c. \qquad S(10)$$

Where $A_\omega$ is the amplitude of the wave at the fundamental incident angular frequency $\omega$. Now, for the if the $A_{3\omega}$ is the amplitude for the third harmonic angular frequency $3\omega$, with the refractive indices of the source and TH are $n_\omega$ and $n_{3\omega}$ Respectively, then the coupled wave equation for THG is given as [6,7]:

$$\frac{dA_\omega}{dz} = \left(\frac{3}{8n_\omega c}\right) i\,\omega\,\chi^{(3)} A_{3\omega}\,|A_\omega|^2\,e^{-i\Delta kz} \qquad S(11a)$$

$$\frac{dA_{3\omega}}{dz} = \left(\frac{3}{8n_{3\omega} c}\right) i\,\omega\,\chi^{(3)}\,|A_\omega|^3\,e^{i\Delta kz} \qquad S(11b)$$

If the source amplitude is constant, then the TH amplitude for a nonlinear medium of height h can be found by integrating Eq. S(11b)

$$A_{3\omega} = \int_0^h \frac{dA_{3\omega}}{dz} dz = \left(\frac{3}{8n_{3\omega}c}\right) i\,\omega\,\chi^{(3)}\,|A_\omega|^3 \int_0^h e^{i\Delta kz} dz$$

$$A_{3\omega} = \left(\frac{3}{8n_{3\omega}c}\right) i\,\omega\,\chi^{(3)}\,|A_\omega|^3\, e^{\frac{i\Delta kh}{2}} \operatorname{sinc}\left(\frac{\Delta kh}{2}\right) \qquad S(12)$$

The intensity of the source beam and the TH beam is:

$$I_\omega = \left(\frac{1}{2}\right) n_\omega c\epsilon_0 |A_\omega|^2 \qquad S(13a)$$

$$I_{3\omega} = \left(\frac{1}{2}\right) n_{3\omega} c\epsilon_0 |A_{3\omega}|^2 = \left[\left(\frac{9\,\pi^2\,\chi^{(3)2}\,h^2}{4\,\epsilon_0^2\,n_\omega^3\,c^2\,\lambda_\omega^2\,n_{3\omega}}\right) \operatorname{sinc}^2\left(\frac{\Delta kh}{2}\right)\right] I_\omega^3 \qquad S(13b)$$

So now I can calculate the internal TH conversion efficiency by the ratio of the transmitted intensity at the TH frequency $3\omega$ to the transmitted intensity at the fundamental frequency $\omega$:

$$\eta_{TH}^{int} = \left(\frac{I_{3\omega}}{I_\omega}\right) \times 100 \qquad S(14)$$

The $\eta_{TH}^{int}$ of equation S(14) is depicted in Figure S1.2 (2), bottom left side, and also the FDTD simulated result is shown in the right bottom side to compare. Although they do not quite match, the similarities in the peak position in terms of the height for TH conversion efficiencies are notable. Although in the same order of magnitude, the TL theory predicts transmission is lower in magnitude, resulting in a lower TH conversion efficiency. However, it also cannot explain why the $\eta_{TH}^{int}$ is so low in magnitude around 1550 nm in height. For this, we calculated the $\chi^{(3)}$ from the TL generated data using the equation using intensity analogue [8]:

$$\chi^{(3)} = \left(\frac{2\,\epsilon_0\,\sqrt{n_{3\omega}\,n_\omega^3}\,c\lambda_\omega}{\sqrt[4]{3}\,\pi\,h\,I_\omega\,T_{Fresnel}\,\operatorname{sinc}\left(\frac{\Delta kh}{2}\right)}\right) \sqrt{\left(\frac{I_{3\omega}}{I_\omega}\right)} \qquad S(15)$$

Where, $T_{Fresnel} = \left(\frac{16\,n_{Substrate,\omega}^2\,n_\omega}{(1+n_{substrate,\omega})^2\,(n_\omega+n_{substrate,\omega})^2}\right) \qquad S(16)$

This intensity $I_{Fresnel} = I_\omega \cdot T_{Fresnel}$ has been plotted along with the TL output, assisted $I_{\omega,TL}$ and $I_{3\omega,TL}$ in Figure S.2.1(b) with the coherent length scaled height of the CaF2 substrate and nonlinear Ge film layer system. Where the coherent length is, by definition [8]:

$$L_{coherent} = \left(\frac{\lambda_\omega}{|\,6n_\omega - 6n_{3\omega}\,|}\right) \qquad S(17)$$

The calculated coherent length is 2467 nm. Usually, the TH power conversion attains a maximum at the coherent length, then drops. However, in FDTD, we do not observe this type of trend, and the TL calculated $\eta_{TH}^{int}$ drop is strictly due to the transmission drop of the fundamental beam. The TH converted intensity fluctuation in Figure S2.1 (b) shows some pattern of quasi phase match condition, where the phase matching condition for the TH generation for the nonlinear medium is due to:

$$\frac{n_\omega \, \omega_\omega}{c} = \frac{n_{3\omega} \, \omega_{3\omega}}{c} \quad \text{so that} \quad \Delta k = 3k(\omega) - K(3\omega) = 0$$

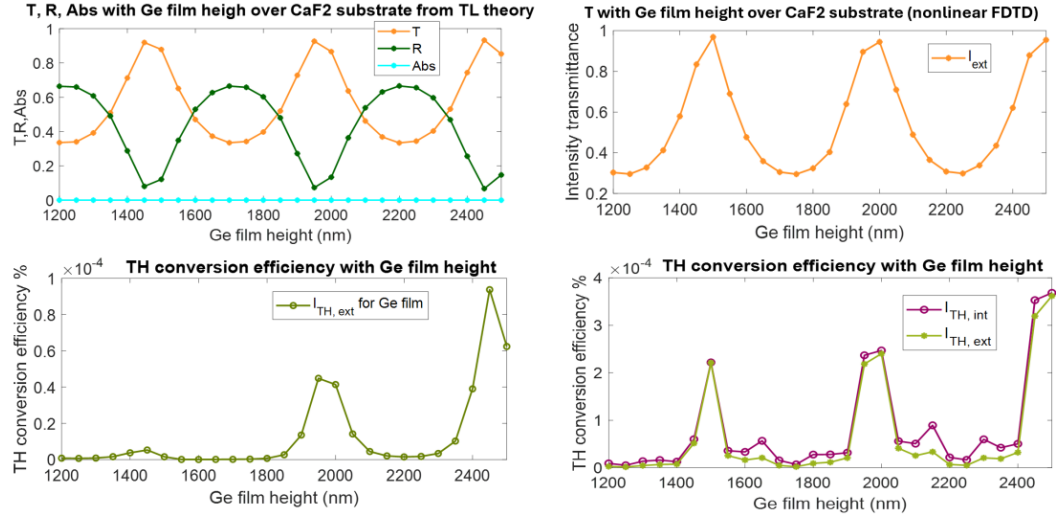

(a)

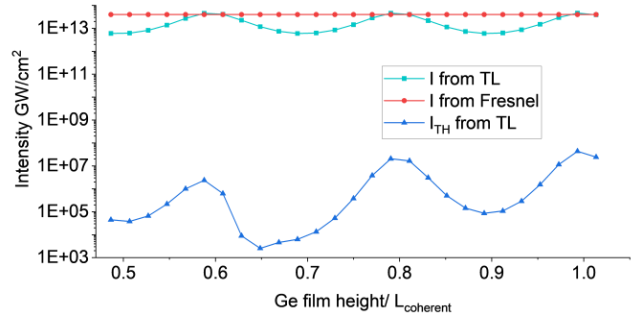

(b)

Fig S2.1. (a) **(Top)** Left side TL theory for calculated T, R, and Abs for CaF2 substrate of 500 nm height and Ge film of increasing height from 1200 nm to 2500 nm. (Right) FDTD simulated result. (Bottom) The comparison of TH conversion efficiency from equation S14 with the transmission output from TL theory (left) and FDTD simulation (right). (b) Transmitted intensity and THG intensity from Equation S17.

### 3.1. Metasurface design unit cell

The individual structural unit of the Ge 2150 nm height metasurface and its TH response to the five order positions will be shown here. In the main text, Figure 3 discusses the TH efficiencies for the desired order position. Here, we demonstrate how one linear polarization order can be suppressed using the asymmetric triangle. Figure S 3.1 (a) The upper right corner of the TE incident TH response is almost two orders of magnitude lower compared to the other three polarization states, and the first order has been suppressed. In Fig. S3.1(b), we observe that the small triangle has a much lower TH efficiency but transfers power to the 1st order under CP incidence. A combination of both the big and small triangles in the same direction suppresses the two orthogonal polarization orders completely and steers the TH power to other polarization states, as we can see in Figure S3.1(c).

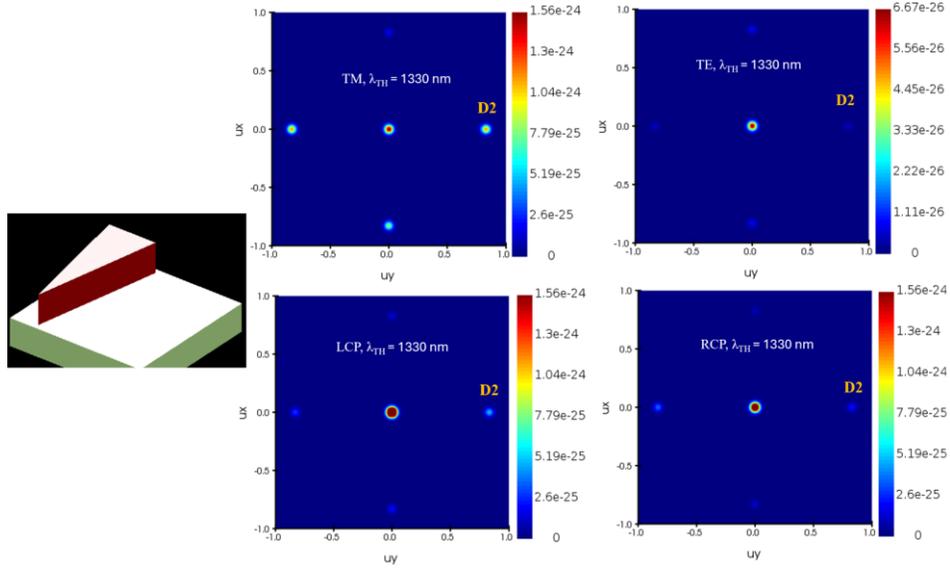

(a)

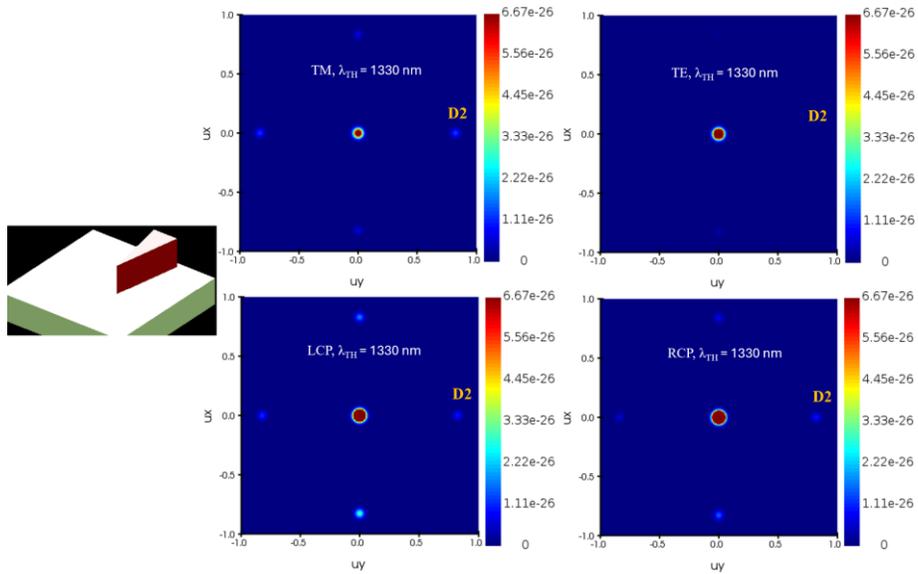

(b)

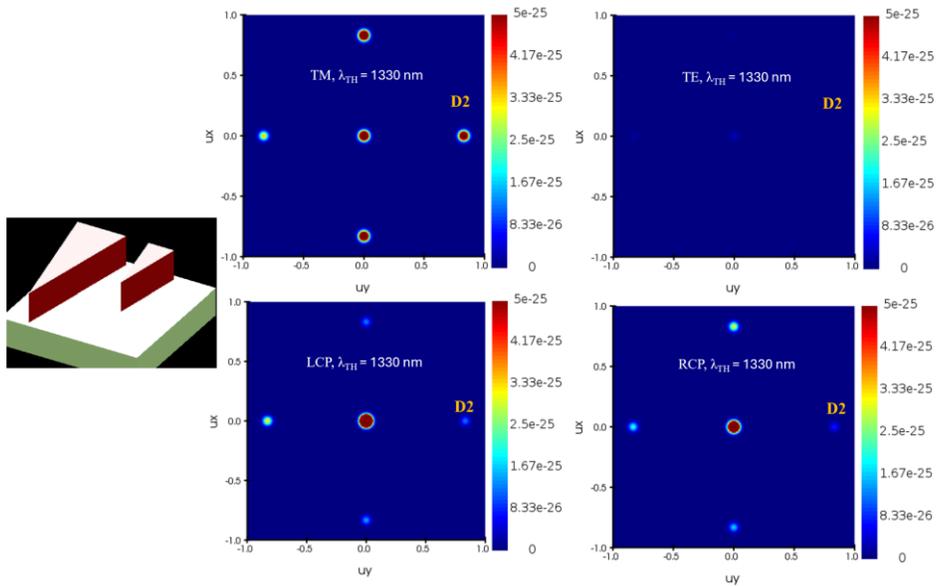

(c)

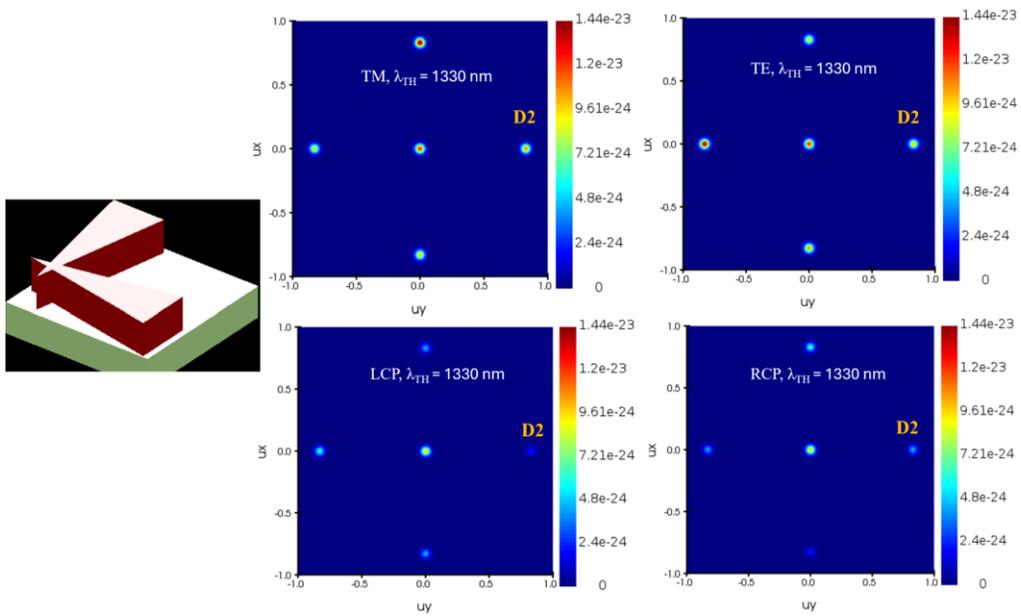

(d)

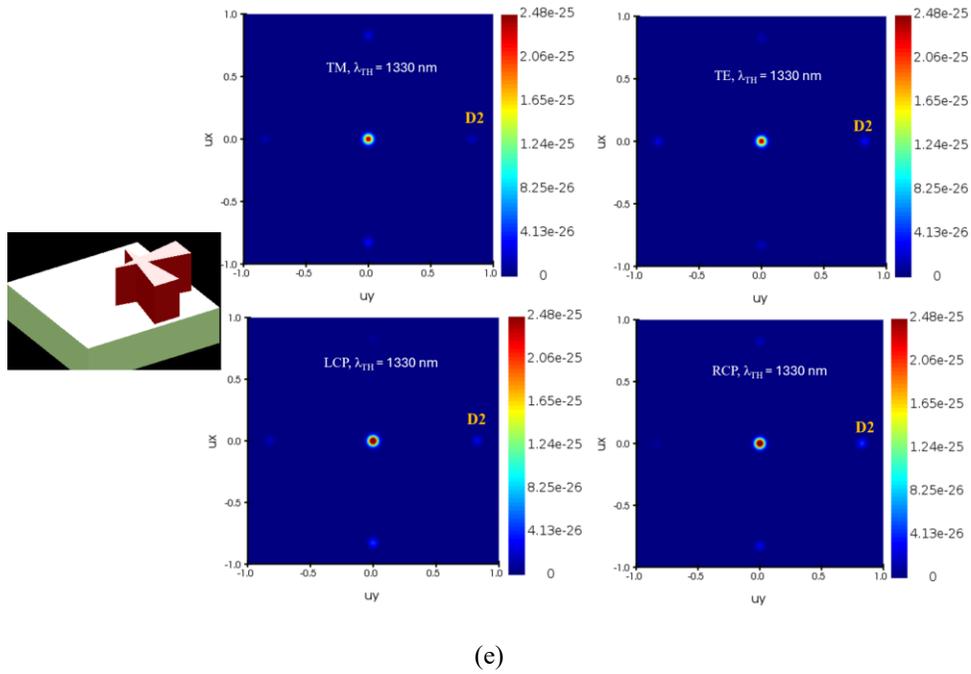

(e)

Fig S3.1: Individual structural unit's TH response to the diffraction order position for the Ge-metasurface. (a) For the single big triangle of length 1478 nm and base 568 nm, align the length towards the x-direction. (b) For the single small triangle of length 820 nm and base 300 nm, align the length with the x-direction. (c) For both single triangles, the length aligns with the x-direction. of length 1478 nm and base 568 nm. (d) Double-cross big triangle. (e) Double-cross small triangle.

$$\theta_{P=1600} = 56.24° \qquad \theta_{P=1950} = 43.01° \qquad \theta_{P=2300} = 35.33°$$

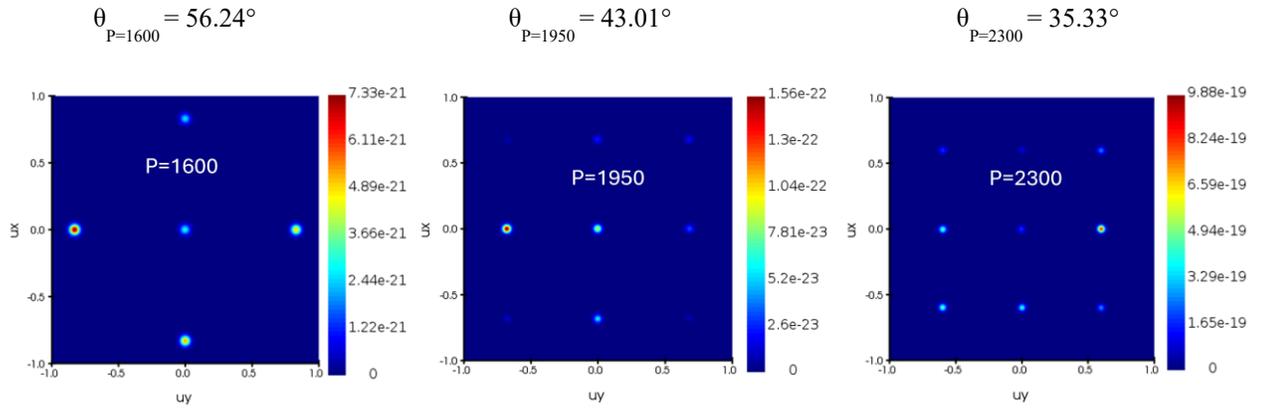

Fig S3.2. (a) Diffraction order angular position with periodicities.

## 4.1. Flipping the diffraction orders choice with incident optical power and anomaly of the Stokes parameter:

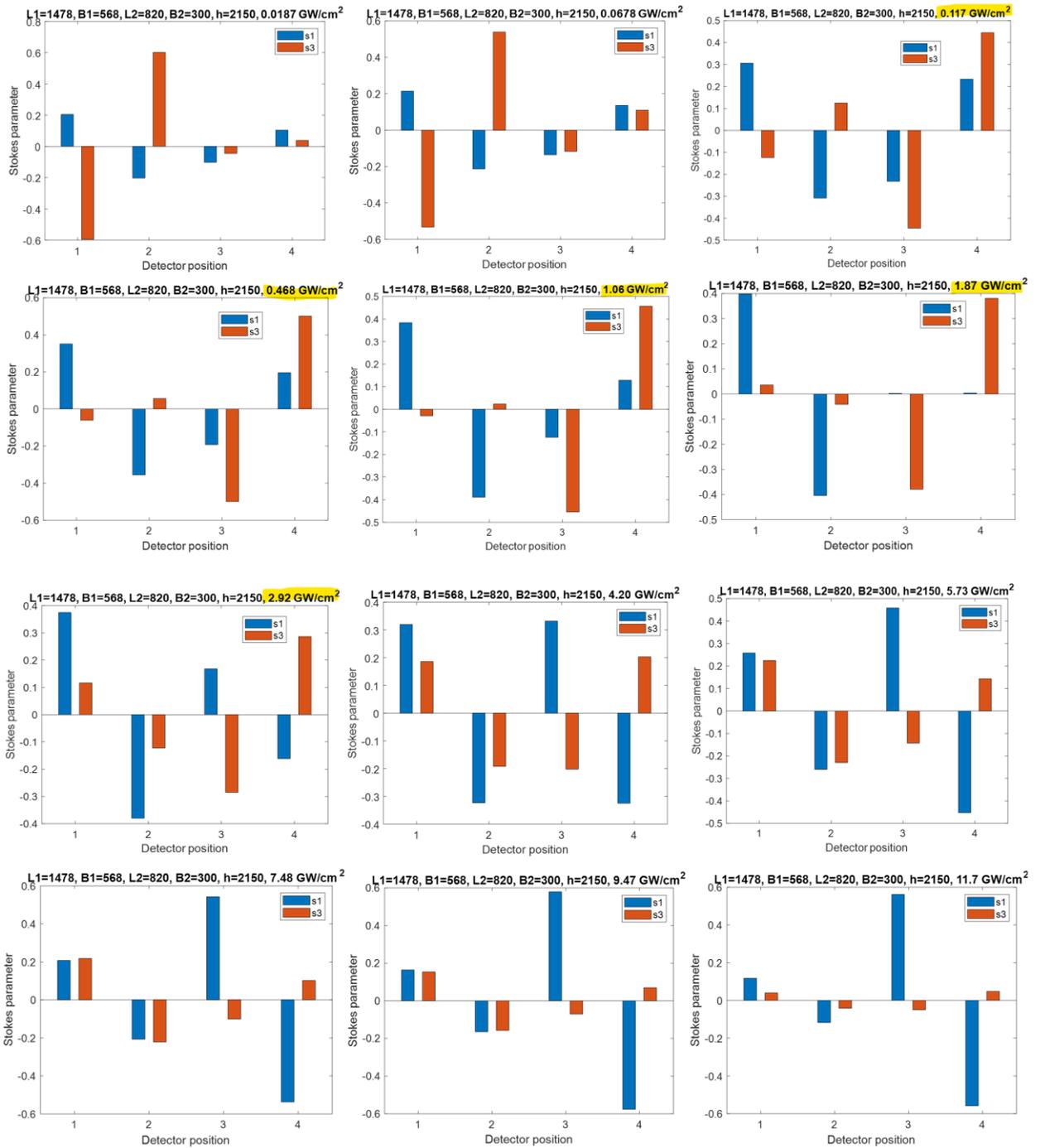

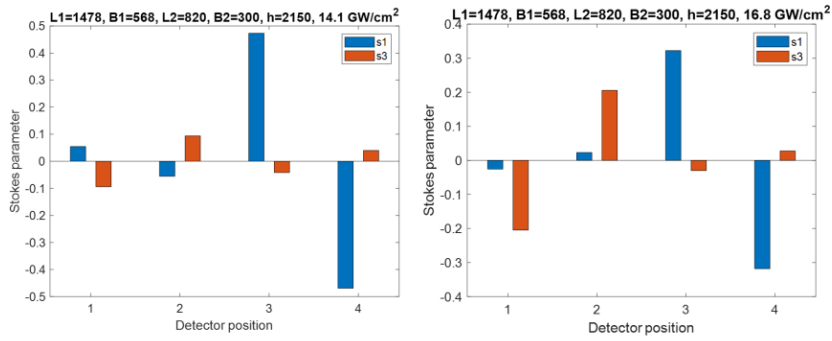

Fig S4.1. Stokes parameter calculation with increasing optical power. The incident intensity of the Stokes parameter is discernible at the desired diffraction order position (in GW/cm²), as indicated by the yellow mark in the top right corner.

## 4.2. Deviation between TH and FH conversion efficiency for the linearly and circularly polarized incident states.

Since this metasurface exhibits the same efficiency for both TE and TM linearly polarized states, we demonstrate that this is also true for LP and CP states. As discussed in the main text, the logarithmic plot of Figure 5. of the TH conversion efficiency rise indicates the quadratic dependence of the source intensity, and in Figure S4.2 (b), the deviation between the LP and the CP states, both very low and high incident intensities, is noticeable, where the Stokes parameter does not go to the desired orders. For low intensity, where third harmonics are just incipient, the CP dominates the LP order position, and for high intensities, where fifth harmonics emerge, the LP dominates at the CP order positions. In the following plot, we see the logarithmic scale of the main text plot in Figure 5, where the area under the yellow region represents the $y = a + bx + cx^2$ behaviors of the TH power raise for both LP and CP incidence, with coefficients almost identical, depending the source power. Interestingly, the TH and FH behaviors differ significantly with this metasurface design at comparatively low intensities.

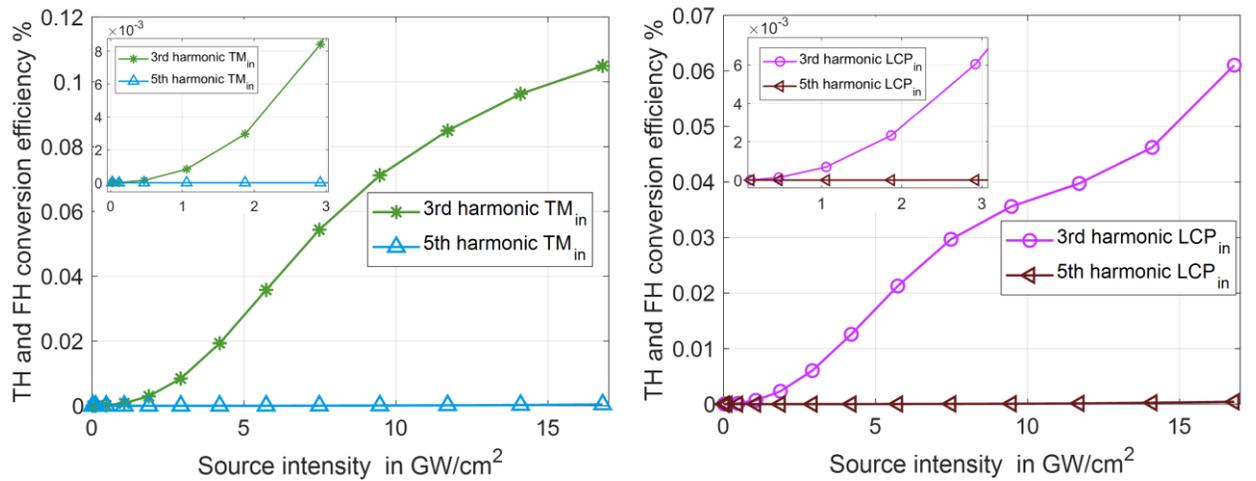

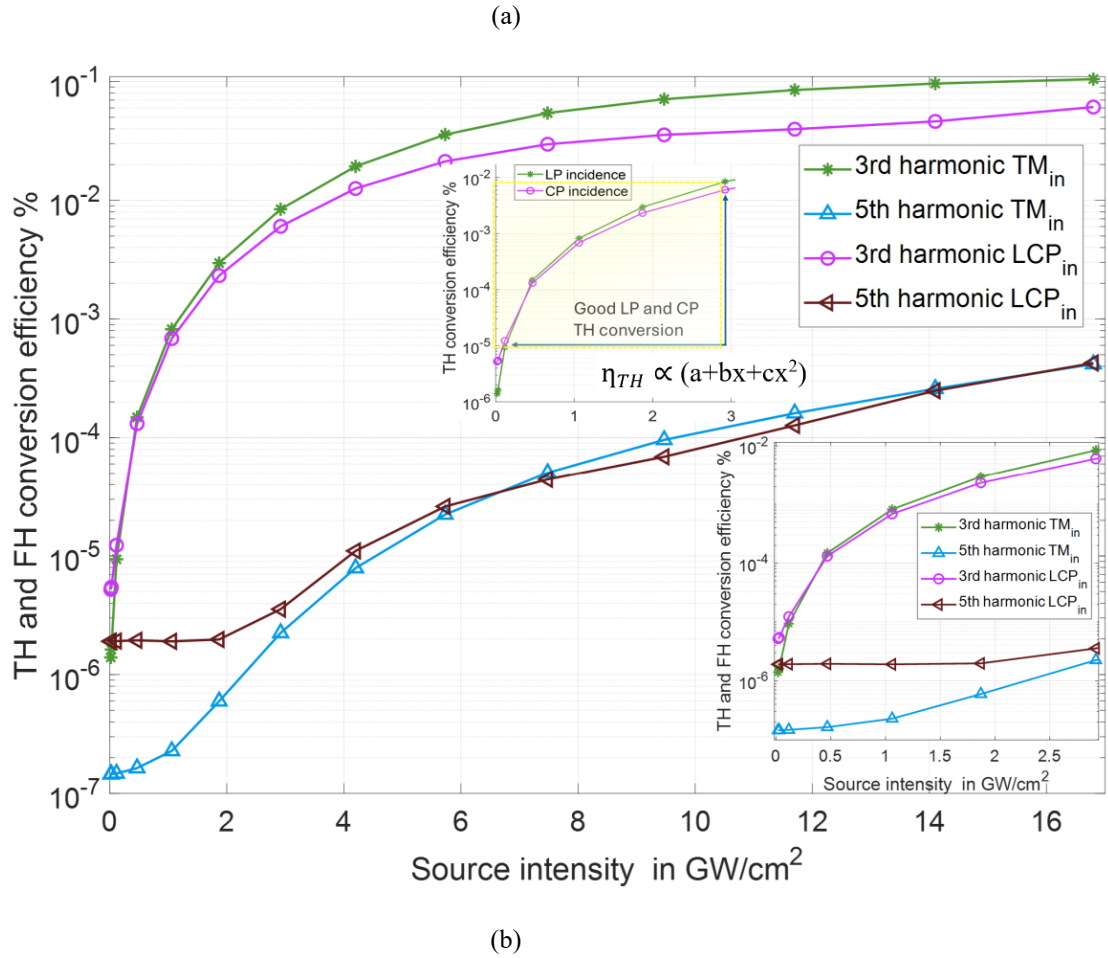

(a)

(b)

Fig S4.2. (a) TH and FH conversion efficiency for LP and CP states. The inset magnifies the region of the source intensity where the Stokes parameters are discernible to the desired diffraction orders. (b) The logarithmic scale of the main text plot in Figure 5 now shows the area under the yellow region, indicating the $y = a + bx + cx^2$ behaviors of TH power rise with incident source power.